\documentclass[preprint,review,11pt,times]{elsarticle}
\usepackage{graphicx,color,soul,amssymb,bm,framed,multicol,multirow,subfig,array,rotating,threeparttable,makecell,textcomp,stfloats,xcolor,lscape,nomencl,hhline,setspace,mathrsfs,indentfirst,balance,enumitem,boldline,booktabs,colortbl,geometry,amsmath}%
\usepackage[numbers]{natbib}
\usepackage[T1]{fontenc}
\usepackage[colorlinks = true,linkcolor=blue, urlcolor=blue, citecolor=blue, anchorcolor = blue]{hyperref}

\journal{Soft Matter}
\begin{document}
\doublespacing
\begin{frontmatter}

\title{Determining hyperelastic properties of the constituents of the mussel byssus system}
\author[address 1]{Yulan Lyu}
\author[address 2]{Yong Pang}
\author[address 2]{Tao Liu\corref{Corresponding author}}
\ead{tao.liu@qmul.ac.uk}
\author[address 1]{Wei Sun}
\cortext[Corresponding author]{Corresponding author}
\address[address 1]{School of Engineering and Materials Science, Queen Mary University of London, London E1 4NS, UK}
\address[address 1]{Faculty of Engineering, University of Nottingham, Nottingham NG7 2RD, UK}

\begin{abstract}
The mussel byssus system, comprising of the adhesive plaque, distal thread, and proximal thread, plays a crucial role in the survival of marine mussels amongst ocean waves. Whilst recent research has explored the stress-strain behaviour of the distal thread and proximal thread through experimental approaches, little attention has been paid to the potential analytical or modelling methods within the current literature. In this work, analytical and finite element (FE) inverse methods were employed for the first time to identify the hyperelastic mechanical properties of both the plaque portion and the proximal thread. The results have demonstrated the feasibility of applied inverse methods in determining the mechanical properties of the constituents of the mussel byssus system, with the residual sum of squares of 0.0004 ($N^2$) and 0.01 ($mm^2$) for the proximal thread and the plaque portion, respectively. By leveraging mechanical and optical tests, this inverse methodology offers a simple and powerful means to anticipate the material properties for different portions of the mussel byssus system, thus providing insights for mimetic applications in engineering and materials design.
\end{abstract}

\end{frontmatter}

\section{Introduction}
\begin{figure*}
 \centering
 \includegraphics[width=1 \linewidth]{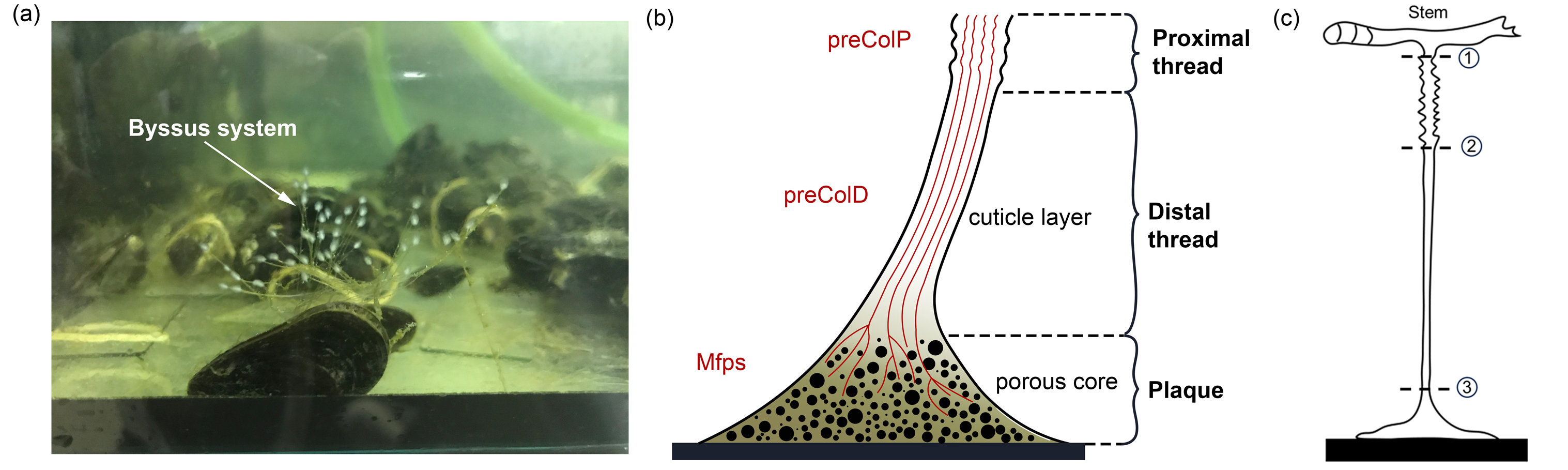}
 \caption{(a) Mussel byssus system showing the threads and plaques, (b) a schematic illustration of a single mussel byssus structure and its protein composition, and (c) a schematic diagram of sample cutting positions: cutting position \textcircled{1} for collecting the whole mussel byssus system samples, \textcircled{1} and \textcircled{3} to collect thread samples, \textcircled{2} and \textcircled{3} for collecting distal portion samples.}
 \label{Fig1}
\end{figure*}

The mussel plaque within the mussel byssus system (Fig. \ref{Fig1}b) can achieve a strong adhesive strength in wet conditions.\cite{lee2011mussel,perepelizin2011thermal,witman1984mussels}. Therefore, scientists have conducted numerous studies on the chemical composition of mussel plaques,\cite{li2019molecular,lin2007adhesion,hwang2010protein} inspiring the development of advanced adhesive materials in medical and engineering fields.\cite{lee2011mussel} For example, six main mussel foot proteins (MFPs) were identified in mussel plaques, which play an essential role in maintaining adhesion at the plaque-substrate interface and establishing connections between threads and plaques.\cite{li2019molecular,lin2007adhesion,hwang2010protein} These MFPs contain high levels of the amino acid 3,4-dihydroxyphenylalanine (DOPA), which forms solid and reversible bonds with various surfaces and is an essential component of adhesive chemistry.\cite{li2019molecular} Meanwhile, studies have also been carried out to investigate the mechanical behaviour of mussel plaques.\cite{allen1976observations,crisp1985adhesion,bell1996mechanical,carrington2004mechanical,burkett2009method,desmond2015dynamics} Desmond et al. (2015) analysed the adhesive performance of mussel plaques subjected to directional tension and summarised three different failure modes: adhesive, interior cohesive, and exterior failure \cite{desmond2015dynamics}. Based on this fundamental research, Pang et al. (2023) further demonstrated that the failure modes of mussel plaques could be changed by the pull angle or the stiffness of the substrate \cite{pang2023quasi}.

Furthermore, as researchers delved deeper into the mussel byssus system, the material properties of the constituents of the mussel threads were also revealed. The mussel thread comprises a distal and a proximal portion. Each portions have different compositions and behave distinct mechanical behaviours that enable mussels to resist dynamic loads and effectively dissipate energy. \cite{bell1996mechanical,holten2008mussel} Specifically, preCol-D and preCol-P, two pre-pepsinized collagens, exhibit a prominent presence in the distal and proximal portions, respectively, potentially leading to different mechanical characteristics. \cite{waite1998peculiar,gosline2002elastic,hagenau2009structural} Gosline and Carrington (2004), revealed that the distal portion of M. californianus mussels can extend approximately half the length of the proximal portion, while demonstrating nearly twice the strength and fifty times the stiffness.\cite{carrington2004mechanical} Furthermore, when subjected to tension, the proximal portion does not exhibit a prominent point of failure, whereas the distal portion experiences four distinct stress regimes: elastic, plateau, hardening, and ultimate failure.\cite{carrington2004mechanical,bell1996mechanical} Additionally, researchers have also performed cyclic loading tests to analyse the self-healing mechanism of mussel threads.\cite{allen1976observations,areyano2022viscoelastic,harrington2009collagen} Reversible sacrificial bonds between proteins and metals are widely recognised to play a crucial role in the self-healing process of mussel threads.\cite{sun2001oxidative,harrington2007holdfast,xu2019metal,zechel2019healing} 

In order to understand the complex mechanics of the mussel byssus system, recent studies have established robust experimental methods to characterise its tensile properties, focusing primarily on the entire thread-plaque system, the mussel thread or solely on its distal portion.\cite{aldred2007tensile,smeathers1979mechanical,carrington2004mechanical} Nevertheless, mechanical properties of proximal portion or plaque portion have yet to be solely studied due to geometry limits and fragility when subjected to tensile loads.\cite{bell1996mechanical} Notably, a fundamental nonlinear spring model was introduced by Waite et al. (2019) to obtain the spring constant and Young’s modulus for the plaque.\cite{mcmeeking2019force} However, these material parameters were obtained by simplifying the stress-strain curve of the combined distal and plaque system into three linear regions.\cite{mcmeeking2019force} Given the similarity between the nonlinear stress-strain behaviour exhibited by mussel threads and the hyperelastic relationship observed in rubber materials,\cite{vaccaro2001yield} the errors in calculating the modulus of the distal and plaque portions by simplifying the overall response to linear are non-negligible. Furthermore, the study have not analyse stress-strain curves for the plaque and distal portion separately.\cite{mcmeeking2019force} Therefore, further investigations on the individual contributions of the distal portion and the plaque to the overall stress-strain response is necessary. 

In the present study, we first investigated the microstructure of three different portions within the mussel byssus system and established correlations between the microstructure and mechanical behaviour. Micro-tensile tests were then performed to obtain tensile force-displacement curves for the distal portion, the entire thread (proximal and distal portions), and the entire plaque-thread system. Meanwhile, the side view of plaque deformation was recorded to verify the inverse Finite Element (FE) approach. Finally, the analytical and FE inverse methods were applied to identify the material properties of the proximal and plaque portions. We validated the application of analytical and FE inverse methods in obtaining material properties for each of the three portions within the mussel byssus system and determined material constitutive models of those portions. The application of the FE inverse approach to obtain the potential plaque bottom surface detachment radius was also discussed in this research. These studies offer demonstrable evidence of the applicability of inverse methods in the analysis of multi-phase biomaterials and provide insight into the mechanical behaviour of each portion of the mussel byssus system for future bio-mimicking. 

\section{Material and methods}
\subsection{Sample preparation procedure}
\subsubsection{Micro-tensile test sample preparation.~~}
Blue mussels (Mytilus edulis) were collected from the coast of Hunstanton in England and kept in a water tank filled with flowing and oxygen-rich seawater at temperatures between 8°C and 10°C. The collected mussels of 3-5 cm in length were tied to 50×50 mm transparent acrylic sheets to allow plaque deposition (the ESI,\dag~ for additional details). Three different sample types were obtained for micro-tensile testing. The first type comprised whole mussel byssus systems, including an adhesive plaque, distal and proximal portions. These were collected by severing mussel threads close to the mussel body using dissecting scissors (Fig. 1c). The second type of samples is mussel threads, including both distal and proximal portions, separated from the adhesive plaques using blades (Fig. 1c). The third type exclusively encompassed the distal portions of mussel threads, collected using dissecting scissors (Fig. 1c).
\subsubsection{Scanning Electron Microscopy (SEM).~~}
The SEM method was conducted to investigate the microstructures of the mussel byssus system. The threads and plaques were cut using dissecting scissors to prepare the SEM samples separately. The samples of the byssal threads and plaques were fixed in a 3.7\% formaldehyde and 2.5\% glutaraldehyde solution for approximately four hours and stored in Milli-Q water for three days at a temperature of 4°C. The samples were first dehydrated in ethanol at increasing concentrations of 50\% (2 times), 70\% (2 times), 90\% (2 times) and 100\% (3 times) and then further dried in Hexamethyldisilane (HDMS) 2 times for 5 mins each.\cite{filippidi2015microscopic} After dehydration and drying, samples were sputter coated with 10 nm thick iridium for 120 s using Quorum Q150V Plus and imaged in JEOL 7100F FEG-SEM using an accelerating voltage of 5 kV. 
\subsection{Micro-tensile tests}
\begin{figure*}
 \centering
 \includegraphics[height=10cm]{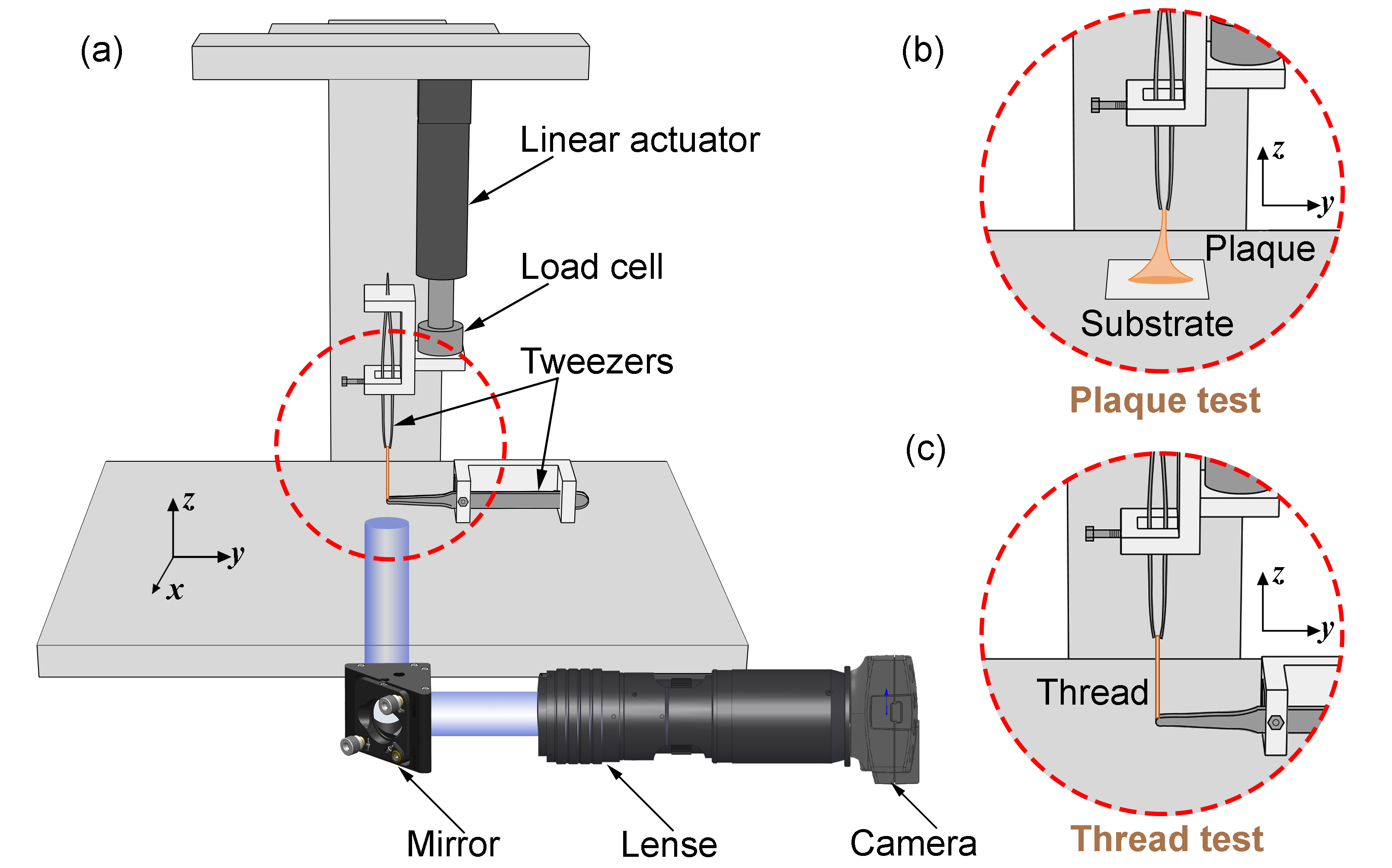}
 \caption{(a) Schematic illustration of a customised micro-tensile experimental setup, (b) focused plaque test setup, and (c) focused thread test setup.}
 \label{Fig2}
\end{figure*}

Two types of micro-tensile tests: plaque tension (Fig. \ref{Fig2}b) and thread tension (distal threads and whole threads) (Fig. \ref{Fig2}c), were conducted on a customised micro-tensile experimental setup (Fig. \ref{Fig2}a). For thread tests, mussel threads were held between two tweezers. The lower tweezer was replaced by a acrylic sheet to which mussel plaques were attached for plaque tests. The upper tweezer is connected to a load cell (Honeywell Model 34, 1000g-10 LBS) to record the loading force. The linear actuator (Thomson MLA11A05) connected to the load cell applies the directional tension to the mussel thread at a slow loading speed of 0.25mm/s to measure the quasi-static tensile responses which are not sensitive to further reduction of loading speed. The linear actuator is powered by a micro-stepping drive (Kollmorgen P5000) connected to the power supply. A monochrome camera (Thorlabs DCC1545M, 1280 x 1024 pixels, 25 fps), lens system (Navitar 12x Macro Zoom) and mirror were positioned on the acrylic plate’s side to record the side view of the plaque deformation process under tension.  
\subsection{Finite Element (FE) forward modelling for mussel plaque}
\begin{figure*}
 \centering
 \includegraphics[height=10cm]{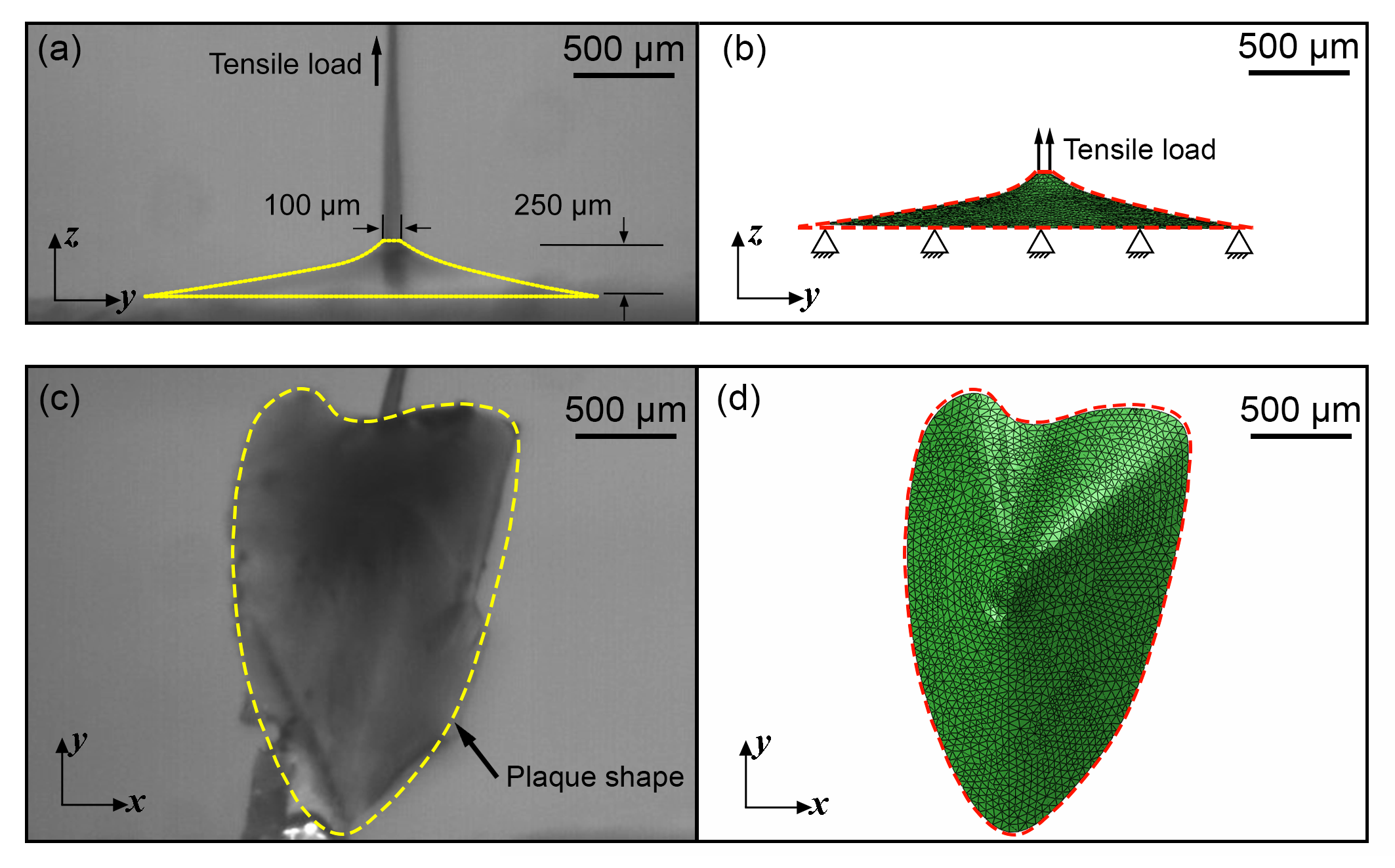}
 \caption{(a) Side view of the mussel plaque, (b) side view of FE model of the mussel plaque, (c) bottom view of the mussel plaque, and (d) bottom view of the FE model of the mussel plaque, all under undeformed configuration.}
 \label{Fig3}
\end{figure*}

A three dimensional (3D) FE model of the mussel plaque was created using ABAQUS/Standard based on the actual measurement from the experiment (Fig. \ref{Fig3}a and c). The diameter of the upper surface and the height of the mussel plaque are approximately 100 µm and 335 µm, respectively (Fig. \ref{Fig3}a). Tensile loads perpendicular to the surface were applied to the upper surface of the plaque using displacement control. The plaque was modelled with 4-node 3D tetrahedral elements in ABAQUS notation (C3D4H elements). The boundary condition of the bottom surface for the plaque model was assumed first to be fully fixed. The Neo Hookean model was applied to present the material properties of mussel plaque. The more detailed equations of this material model can be seen in Section 3.1. The material properties of the mussel plaque model were iterated in the inverse optimisation algorithm until the tensile force-displacement curve and the deformed shapes derived from the FE model synchronised with the results obtained experimentally.
\subsection{Fundamental of inverse approaches}
It was assumed that within a composite material comprising a series of sub-materials, there exists an unidentified sub-material with unknown material parameters. Therefore, the inverse method can be applied to identify the material properties of the unknown sub-material by simplifying the composite as a series spring system (Fig. \ref{Fig4}a). The total number of sub-materials was assumed to be \(p\), where the material parameters of the sub-materials with number \(p-1\) are known (Fig. \ref{Fig4}a). Therefore, three main inputs in the analytical inverse approach were as follows: 
\begin{figure*}
 \centering
 \includegraphics[width=1 \linewidth]{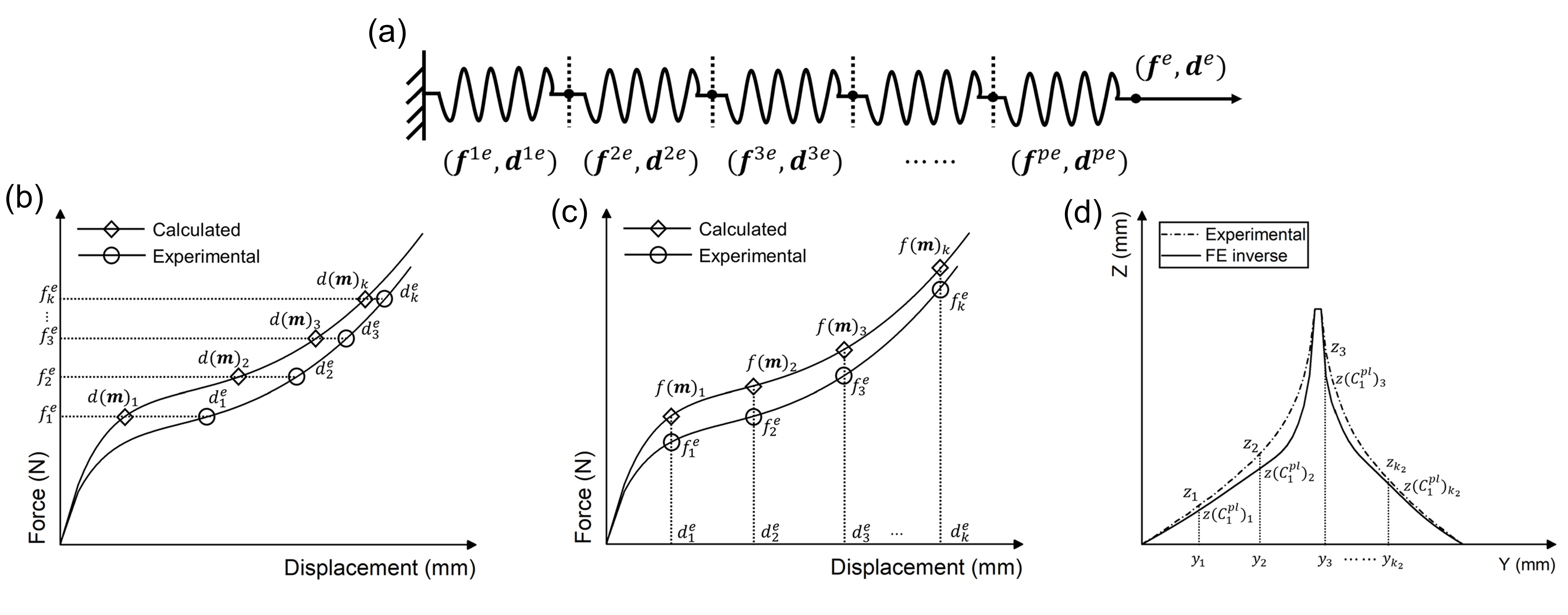}
 \caption{(a) Schematic diagram of the simplified spring system, (b) schematic representation of the principle of the inverse method using displacement as a metric, (c) schematic representation of the principle of the inverse method using force as a metric, and (d) schematic of the deformed shape optimisation approach.}
 \label{Fig4}
\end{figure*}
\begin{enumerate}
    \item The experimental force-displacement data of the overall spring system $\left(\textbf{\textit{f}}^e, \textbf{\textit{d}}^e\right)$, where $\textbf{\textit{f}}^e=\left[f_1^e, f_2^e, f_3^e, \cdots, f_k^e\right]^T$, $\textbf{\textit{d}}^e=\left[d_1^e, d_2^e, d_3^e, \cdots, d_k^e\right]^T$, \(k\) is the number of data points obtained from the force-displacement curves (Fig. \ref{Fig4}b and c).
    \item Material parameters of known sub-materials \(\boldsymbol{\omega}_i\), where \(i=1, 2, 3, \cdots, p-1.\)
    \item Guessing material parameters of the unknown sub-material $\textbf{\textit{m}}=\left[m_1, m_2, m_3, \cdots, m_q\right]^T$, where \(q\) is the number of material parameters for the unknown sub-material.    
\end{enumerate}

As the spring system is compounded in series, the tensile force of the overall spring system should be equal to the tensile force of any sub-material, and the tensile displacement of the overall spring system should equal to the sum of the elongations of all sub-materials:
\[\textbf{\textit{f}}^e=\textbf{\textit{f}}^{1 e}=\textbf{\textit{f}}^{2 e}=\ldots=\textbf{\textit{f}}^{p e}\]
\begin{equation}
 \textbf{\textit{d}}^e=\textbf{\textit{d}}^{1 e}+\textbf{\textit{d}}^{2 e}+\cdots+\textbf{\textit{d}}^{p e}
\end{equation}

Because the material parameters \(\boldsymbol{\omega}_i\), material constitutive equations and sizes of the \(p-1\) sub-materials were assumed to be known, the elongations of these sub-materials can be calculated based on eqn (1). Meanwhile, the elongation of the unknown sub-material \(\textbf{\textit{d}}^u(\textbf{\textit{m}})\) can also be calculated using the guessing material parameters \(\textbf{\textit{m}}\), assumed material constitutive equation, measured sizes and the force of unknown sub-materials \(\textbf{\textit{f}}^u(\textbf{\textit{m}})=\textbf{\textit{f}}^e\). Notably, \(\textbf{\textit{d}}^u(\textbf{\textit{m}})\) and \(\textbf{\textit{f}}^u(\textbf{\textit{m}})\) denote their values are functions of \(\textbf{\textit{m}}\), and the rule applies to subsequent arguments containing \(\textbf{\textit{m}}\) in parentheses. 

Therefore, the displacement of the overall spring system from the analytical inverse approach \(\textbf{\textit{d}}(\textbf{\textit{m}})\) will be:
\begin{equation}
\textbf{\textit{d}}(\textbf{\textit{m}})=\textbf{\textit{d}}^{1}+\textbf{\textit{d}}^{2}+\cdots+\textbf{\textit{d}}^{p-1}+\textbf{\textit{d}}^u(\textbf{\textit{m}})
\end{equation}
where \(\textbf{\textit{d}}^{1}, \textbf{\textit{d}}^{2}, \cdots, \textbf{\textit{d}}^{p-1}\) are the elongation vectors for known sub-materials calculated from the known material parameters \(\boldsymbol{\omega}_i\), material constitutive models, measured sizes and experimental forces.

Thus, the residual vector \(\textbf{\textit{F}}(\textbf{\textit{m}})\) based on the guessing material parameters \(\textbf{\textit{m}}\) can be defined as (Fig. \ref{Fig4}b):
\begin{equation}
\textbf{\textit{F}}(\textbf{\textit{m}})=\textbf{\textit{d}}^e-\textbf{\textit{d}}(\textbf{\textit{m}})=\left[F_1(\textbf{\textit{m}}), F_2(\textbf{\textit{m}}), F_3(\textbf{\textit{m}}), \ldots, F_k(\textbf{\textit{m}})\right]^T
\end{equation}

If the force is used as the optimisation metrics, the residual vector \(\textbf{\textit{F}}(\textbf{\textit{m}})\) can be determined as (Fig. 4c):
\begin{equation}
\textbf{\textit{F}}(\textbf{\textit{m}})=\textbf{\textit{f}}^e-\textbf{\textit{f}}(\textbf{\textit{m}})
\end{equation}
where \(\textbf{\textit{f}}(\textbf{\textit{m}})\) is the force vector of the overall spring system based on the guessing material parameters \(\textbf{\textit{m}}\) from the analytical inverse approach. 

The optimisation process is achieved by applying the nonlinear least squares estimation function \textbf{lsqnonlin} within the Matlab Optimisation Toolbox.\cite{guide1998mathworks} This method effectively addresses curve-fitting problems characterised by using the following form of nonlinear least-squares:\cite{guide1998mathworks}
\begin{equation}
\min _\textbf{\textit{m}}\|\textbf{\textit{F}}(\textbf{\textit{m}})\|_2^2=\min _\textbf{\textit{m}}\left(F_1(\textbf{\textit{m}})^2+F_2(\textbf{\textit{m}})^2+F_3(\textbf{\textit{m}})^2+\cdots+F_k(\textbf{\textit{m}})^2\right)
\end{equation}

The lower and upper bounds \textbf{lb} and \textbf{ub} were set to limit the value ranges of material parameters \textbf{\textit{m}} and thus accelerate the inverse approach. The functional tolerance \(\theta\) is initially set to 0.01 in the ‘options’ and can be changed to the desired degree of precision.\cite{guide1998mathworks} When the final change in the sum of squares relative to its initial value is less than the value of the function tolerance, the \textbf{lsqnonlin} will stop and the optimised material parameters \(\textbf{\textit{m}}^*\) will be obtained (Fig. 5):
\begin{equation}
\min _\textbf{\textit{m}}\|\textbf{\textit{F}}(\textbf{\textit{m}})\|_2^2 \leq \theta
\end{equation}

However, if the shape of the material is irregular, the analytical inverse approach is no longer applicable, and the FE-based inverse approach is needed. The ".inp” file for ABAQUS was first written, including the material geometry, known and guessed material parameters, material constitutive models and boundary conditions. A post-processing Python script was then created to extract force-displacement results based on guessing material parameters from the ABAQUS “.odb” file. The remaining optimisation process is consistent with the analytical inverse method by applying \textbf{lsqnonlin}, meaning that when eqn (6) is satisfied, the optimised material parameters \(\textbf{\textit{m}}^*\) will be obtained (Fig. 5).
\begin{figure}[h]
\centering
  \includegraphics[height=10cm]{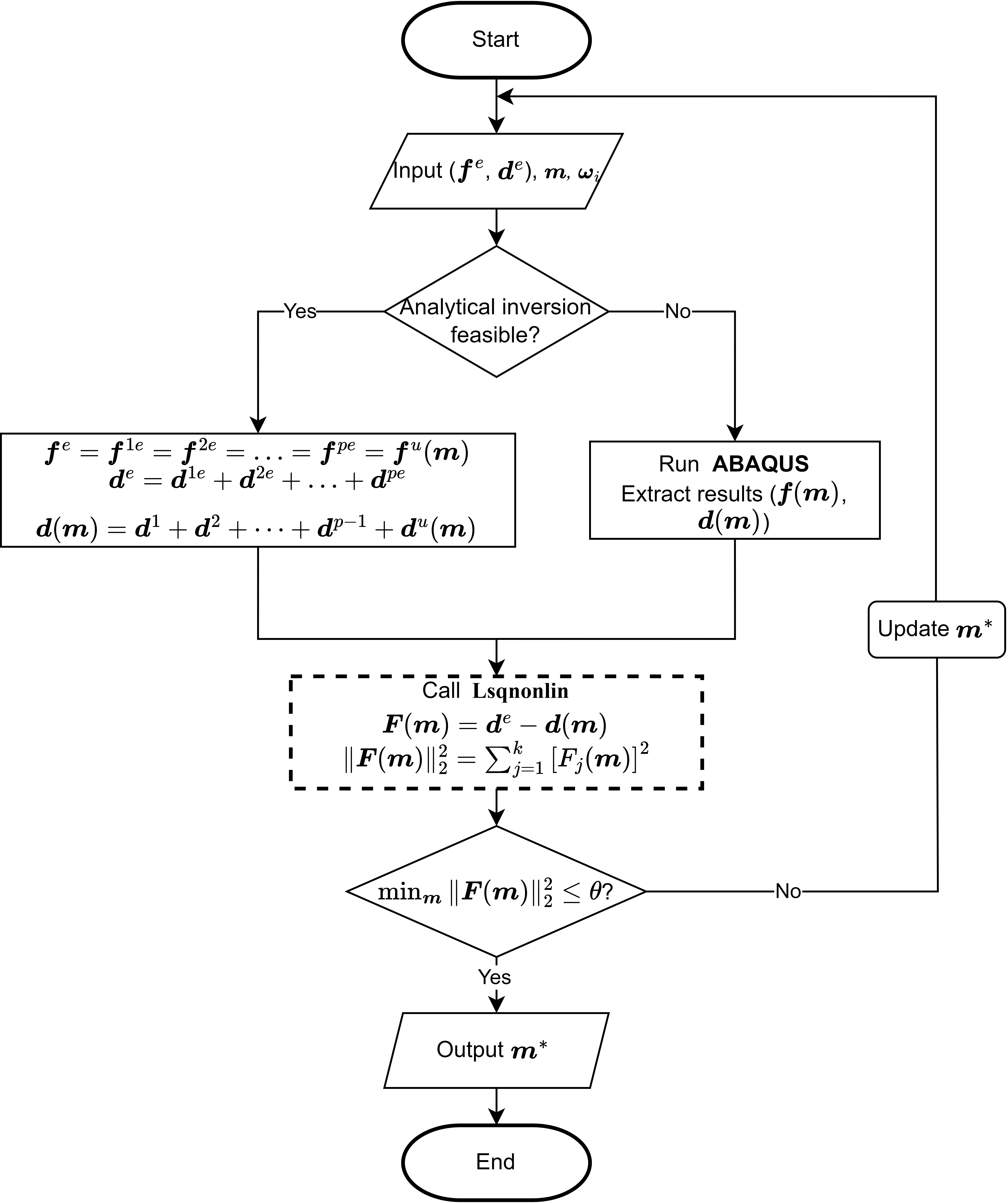}
  \caption{Flow chart of both analytical inverse and FE-based inverse procedures.}
  \label{Fig5}
\end{figure}
\section{Mussel byssus system material constitutive models and inverse procedure}
\subsection{Hyperelastic models}
Based on the characteristics of the stress-strain curves of the distal and proximal threads and the force-displacement curves of the entire byssus system in the previous literature,\cite{carrington2004mechanical,bell1996mechanical,desmond2015dynamics,waite2002elastomeric} different hyperelastic materials constitutive models were assumed to present plaques, distal threads and proximal threads, respectively. The incompressible Ogden’s model was applied to the distal threads because the shape of the strain-stress curves for the distal threads has characteristics similar to those of the Ogden's model curve. Similarly, the constitutive model of both the proximal portion and the plaque was assumed to be the incompressible Neo-Hookean material model due to the resemblance of the curve shape. 

The strain energy potential \(W\) of the three-order Ogden model is expressed in terms of principal stretches as follows:\cite{ogden1972large} 
\begin{equation}
W=\sum_{\beta=1}^N \frac{2 \mu_\beta}{\alpha_\beta^2}\left(\bar{\lambda}_1^{\alpha_\beta}+\bar{\lambda}_2^{\alpha_\beta}+\bar{\lambda}_3^{\alpha_\beta}-3\right)
\end{equation}
where \(N\) is the order of strain energy potential (\(N=3\)), \(\mu_\beta\) and \(\alpha_\beta\) \((\beta=1, 2, 3)\) are six material constants. \(\bar{\lambda}_1\), \(\bar{\lambda}_2\) and \(\bar{\lambda}_3\) are the distortional principal stretches which can be calculated using principal stretches \(\bar{\lambda}_\eta=J^{-\frac{1}{3}} \cdot \lambda_\eta(\eta=1,2,3)\), \(J\) is the total volume ratio. 

Meanwhile, the strain energy potential of the incompressible Neo-Hookean material model is given below:\cite{rivlin1948large}
\begin{equation}
W=C_1\left(\bar{\lambda}_1{ }^2+\bar{\lambda}_2{ }^2+\bar{\lambda}_3{ }^2-3\right)
\end{equation}
where \(C_1\) is a material constant. 

Tensile tests were assumed to be uniaxial. Therefore, the uniaxial stress \(\sigma\) and stretch \(\lambda\) for Ogden’s model can be expressed as
\begin{equation}
\sigma=\sum_{\beta=1}^3 \frac{2 \mu_\beta}{\alpha_\beta}\left[\lambda^{\alpha_\beta}-\left(\frac{1}{\sqrt{\lambda}}\right)^{\alpha_\beta}\right]
\end{equation}

The uniaxial stress \(\sigma\) and stretch \(\lambda\) for Neo-Hookean hyperelastic model are defined as shown below:
\begin{equation}
\sigma=2 C_1\left(\lambda-\frac{1}{\lambda^2}\right)
\end{equation}
\subsection{Application of inverse approaches to mussel byssus system}
\subsubsection{Problem definition.~~}
The shorter length of the proximal portion compared to the distal portion of threads and the irregular shape of mussel plaques increase the difficulty of the micro-tensile tests. Therefore, analytical and FE inverse approaches were employed to obtain their material properties.
\subsubsection{Analytical inverse approach for the proximal thread.~~}
The three portions of the mussel byssus system were simplified as a spring system in series. Tensile force and displacement data for the distal portion \((\textbf{\textit{f}}^e_d, \textbf{\textit{d}}^e_d)\) was obtained from the micro-tensile tests for the distal portion, and the stress-strain data \((\boldsymbol{\sigma}_d, \boldsymbol{\varepsilon}_d)\) can be further calculated by
\[\boldsymbol{\sigma}^e_d=\frac{\textbf{\textit{f}}^e_d}{\pi r_d^2}\]
\begin{equation}
\boldsymbol{\varepsilon}^e_d=\frac{\textbf{\textit{d}}^e_d}{l_d}
\end{equation}
where \(r_d\) and \(l_d\) are the radius and length of the distal portion before tension.

Based on the experiment data and the Hyperelastic Ogden model (eqn (9)), six material parameters of the distal portion (\(\mu_\beta\), \(\alpha_\beta\)), \((\beta=1, 2, 3)\) can be determined using a simple analytical or FE-based inverse approach shown in Fig. \ref{Fig5} and were used as input for the further analytical inverse approach. 

Next, the tensile force and displacement data of the whole thread (distal and proximal portions) \((\textbf{\textit{f}}^e_t,\textbf{\textit{d}}^e_t)\) was also obtained from micro-tensile tests for the whole thread. Based on eqn (1), the following euqations can be obtained: 
\[\textbf{\textit{f}}^e_t=\textbf{\textit{f}}^{e}_p=\textbf{\textit{f}}^{e}_d\]
\begin{equation}
 \textbf{\textit{d}}^e_t=\textbf{\textit{d}}^{e}_p+\textbf{\textit{d}}^{e}_d
\end{equation}
where \(\textbf{\textit{f}}^{e}_p=\left[(f_{p}^e)_1, (f_{p}^e)_2, \cdots, (f_{p}^e)_k\right]^T\) and \(\textbf{\textit{f}}^{e}_t=\left[(f_{t}^e)_1, (f_{t}^e)_2, \cdots (f_{t}^e)_k\right]^T\) are the tensile force vectors of the proximal portion and the distal portion, respectively. \(\textbf{\textit{d}}^{e}_p=\left[(d_{p}^e)_1, (d_{p}^e)_2, \cdots, (d_{p}^e)_k\right]^T\) and \(\textbf{\textit{d}}^{e}_t=\left[(d_{t}^e)_1, (d_{t}^e)_2, \cdots, (d_{t}^e)_k\right]^T\) are the elongation vectors of the proximal portion and the distal portion, respectively.  

Therefore, the elongations of the distal portion can be calculated based on eqn (9), (11) and (12), and the six material parameters obtained from the previous distal portion tensile tests:
\begin{equation}
\frac{\left(f^e_t\right)_j}{\pi r_d^2}=\sum_{\beta=1}^3 \frac{2 \mu_\beta}{\alpha_\beta}\left[\left(\frac{\left(d^e_d\right)_j+l_d}{l_d}\right)^{\alpha_\beta}-\left(\frac{1}{\sqrt{\frac{\left(d^e_d\right)_j+l_d}{l_d}}}\right)^{\alpha_\beta}\right]
\end{equation}
where \(j=1, 2, 3, \cdots, k \)

Meanwhile, the elongation of the proximal portion \(\textbf{\textit{d}}(C^p_1)_p\) based on the given material constant \(C^p_1\) can be calculated in the analytical inverse approach based on eqn (10) and (12):
\begin{equation}
\left(d\left(C^p_1\right)_p\right)_j=l_p\left(\frac{\left(f_t^e\right)_j}{2\pi C^p_1 r_p^2}+\frac{1}{\left(\frac{\left(d\left(C^p_1\right)_p\right)_j+l_p}{l_p}\right)^2}-1\right)
\end{equation}
where \(r_p\) and \(l_p\) are the radius and length of the proximal portion before tension. \(\left(d\left(C^p_1\right)_p\right)_j\) is one data point within \(\textbf{\textit{d}}(C^p_1)_p\) and a function of \(C^p_1\). (The rule also applies to subsequent arguments containing \(C^p_1\) in parentheses)

As a result, the tensile displacement of the whole thread based on the given material constant \(C^p_1\) can be expressed as
\begin{equation}
 \textbf{\textit{d}}(C^p_1)_t=\textbf{\textit{d}}(C^p_1)_p+\textbf{\textit{d}}^{e}_d
\end{equation}

The residual vector \(\textbf{\textit{F}}(C^p_1)\) based on the given material constant \(C^p_1\) then can be defined as
\begin{equation}
\textbf{\textit{F}}(C^p_1)=\textbf{\textit{d}}^e_t-\textbf{\textit{d}}(C^p_1)_t=\left[F_1(C^p_1), F_2(C^p_1), F_3(C^p_1), \ldots, F_k(C^p_1)\right]^T
\end{equation}

If the tensile force was used as metrics, the residual vector \(\textbf{\textit{F}}(C^p_1)\) can be expressed as
\begin{equation}
\textbf{\textit{F}}(C^p_1)=\textbf{\textit{f}}^e_t-\textbf{\textit{f}}(C^p_1)_t
\end{equation}

\textbf{lsqnonlin} was applied for parameter optimisation as Section 2.4, and the functional tolerance \(\theta\) was initially set as 0.01. \textbf{lsqnonlin} will stop when the following convergence criterion is satisfied:
\begin{equation}
\min _{C^p_1}\left\|\textbf{\textit{F}}\left(C^p_1\right)\right\|_2^2 \leq \theta
\end{equation}
\subsubsection{FE-based inverse approach for the mussel plaque.~~}
The FE inverse approach was employed as the irregular shape of mussel plaques makes the above analytical inverse approach not applicable. Micro-tensile tests were first performed on the combined distal and plaque portions because it was difficult to perform experiments on the plaques themselves. Based on eqn (1), the force of the mussel plaque \(\textbf{\textit{f}}_{p l}^e\) can be obtained from the experiment:
\begin{equation}
\textbf{\textit{f}}_{p l}^e=\textbf{\textit{f}}_d^e=\textbf{\textit{f}}_{d+p l}^e
\end{equation}
where \(\textbf{\textit{f}}_d^e\) and \(\textbf{\textit{f}}_{d+p l}^e \) are the tensile force vectors of the distal portion and distal-plaque system, respectively. 

Meanwhile, the tensile displacement of the upper surface of the mussel plaque \(\textbf{\textit{d}}_{p l}^e\) was also measured from the side-view video (Fig. 11a) to generate the experimental force-displacement data \((\textbf{\textit{f}}_{p l}^e, \textbf{\textit{d}}_{p l}^e)\) as input in the FE inverse approach. The guessing material parameter \(C^{p l}_1\) of the mussel plaque was then written on the ABAQUS CAE Interface and updated in \textbf{lsqnonlin}. A post-processing Python script was created to extract the force-displacement data \((\textbf{\textit{f}}(C^{p l}_1)_{p l}, \textbf{\textit{d}}(C^{p l}_1)_{p l})\) based on the guessing material paramter \(C^{p l}_1\) from the ABAQUS odb file. 

The residual vector \(\textbf{\textit{F}}(C^{p l}_1)\) and the sum of squared residuals from the force-displacement data based on the guessing material paramter \(C^{p l}_1\) can be defined as:
\[\textbf{\textit{F}}(C^{p l}_1)=\textbf{\textit{d}}^e_{p l}-\textbf{\textit{d}}(C^{p l}_1)_{p l}=\left[F_1(C^{p l}_1), F_2(C^{p l}_1), \ldots, F_k(C^{p l}_1)\right]^T\]
\begin{equation}
\left\|\textbf{\textit{F}}\left(C^{p l}_1\right)\right\|_2^2=\sum_{j=1}^k\left[F_j\left(C^{p l}_1\right)\right]^2
\end{equation}

To further optimise the FE inverse approach, the deformed side views of the mussel plaque measured from the experiment were compared with the FE results. It was assumed that the deformed shapes with a total number of \(k_1\) were analysed. For each deformed shape, the leftmost node was set as (0,0) (Fig. 3a), and nodes with a total number of \(k_2\) were selected on each shape. Graph Grabber v 2.0.1 software was used to obtain the coordinates \((\textbf{\textit{y}},\textbf{\textit{z}})\) of selected nodes (\url{https://www.quintessa.org/software}). Meanwile, the coordinates \((\textbf{\textit{y}},\textbf{\textit{z}}(C^{p l}_1))\) of corresponding nodes based on the guessing material paramter \(C^{p l}_1\) were exported from the FE model using a post-processing Python script. 

The residual vector \(\textbf{\textit{F}}^s(C^{p l}_1)\) of each deformed shape can be expressed as (Fig. \ref{Fig4}d)
\begin{equation}
\textbf{\textit{F}}^s(C^{p l}_1)=\textbf{\textit{z}}-\textbf{\textit{z}}(C^{p l}_1)=\left[F^s_1(C^{p l}_1), F^s_2(C^{p l}_1), \ldots, F^s_{k_2}(C^{p l}_1)\right]^T
\end{equation}

Thus, the sum of squared residuals for all the deformed shapes were defined as:
\begin{equation}
\left\|\textbf{\textit{F}}^s\left(C^{p l}_1\right)\right\|_2^2=\sum_{h=1}^{k_1} \sum_{g=1}^{k_2}\left[F_g^s\left(C^{p l}_1\right)_h\right]^2
\end{equation}

The total sum of squared residuals \(R(C^{p l}_1)\) for both force-displacement curves and deformed shapes is defined below: 
\begin{equation}
R(C^{p l}_1)=W_1 \cdot \left\|\textbf{\textit{F}}\left(C^{p l}_1\right)\right\|_2^2+W_2 \cdot \left\|\textbf{\textit{F}}^s\left(C^{p l}_1\right)\right\|_2^2
\end{equation}
where \(W_1\) and \(W_2\) are the non-negative weights, \(W_1, W_2\in[0,1]\),\\ \(W_1+W_2=1\). \(W_1\) and \(W_2\) were set to 0.8 and 0.2, respectively, indicating that the accuracy of the force-displacement data is more critical in this research. Their values can be adjusted according to study interest.\cite{Schnur,sun2022overview}

\textbf{lsqnonlin} was applied for parameter optimisation as Section 2.4, and the functional tolerance \(\theta\) was initially set as 0.01. The value of \(\theta\) can be adjusted based on the optimisation condition. \textbf{lsqnonlin} will stop when the following convergence criteria are satisfied:
\begin{equation}
\min _{C^{p l}_1}[R(C^{p l}_1)] \leq \theta
\end{equation}

\begin{figure*}
 \centering
 \includegraphics[width=1 \linewidth]{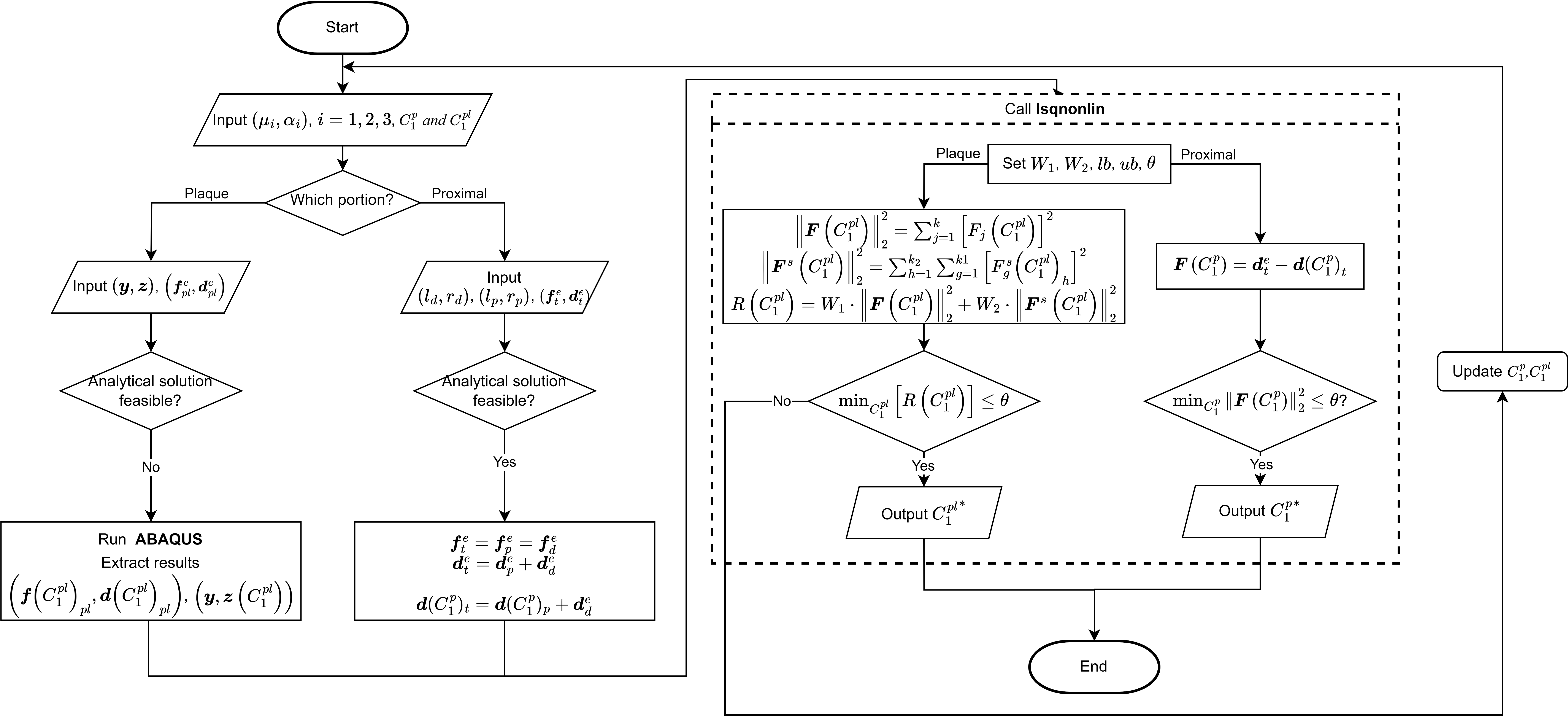}
 \caption{Flow chart of determining the hyperelastic properties of the plaque and the proximal portion by using FE-based inverse approach and analytical inverse approach, respecticvely.}
 \label{Fig6}
\end{figure*}

The analytical inverse and FE-based inverse procedures specifically for the proximal portion and the plaque portion are summarised in Fig. \ref{Fig6}. 
\section{Results}
\subsection{Microstructures of the mussel byssus system}
\begin{figure*}
 \centering
 \includegraphics[width=1 \linewidth]{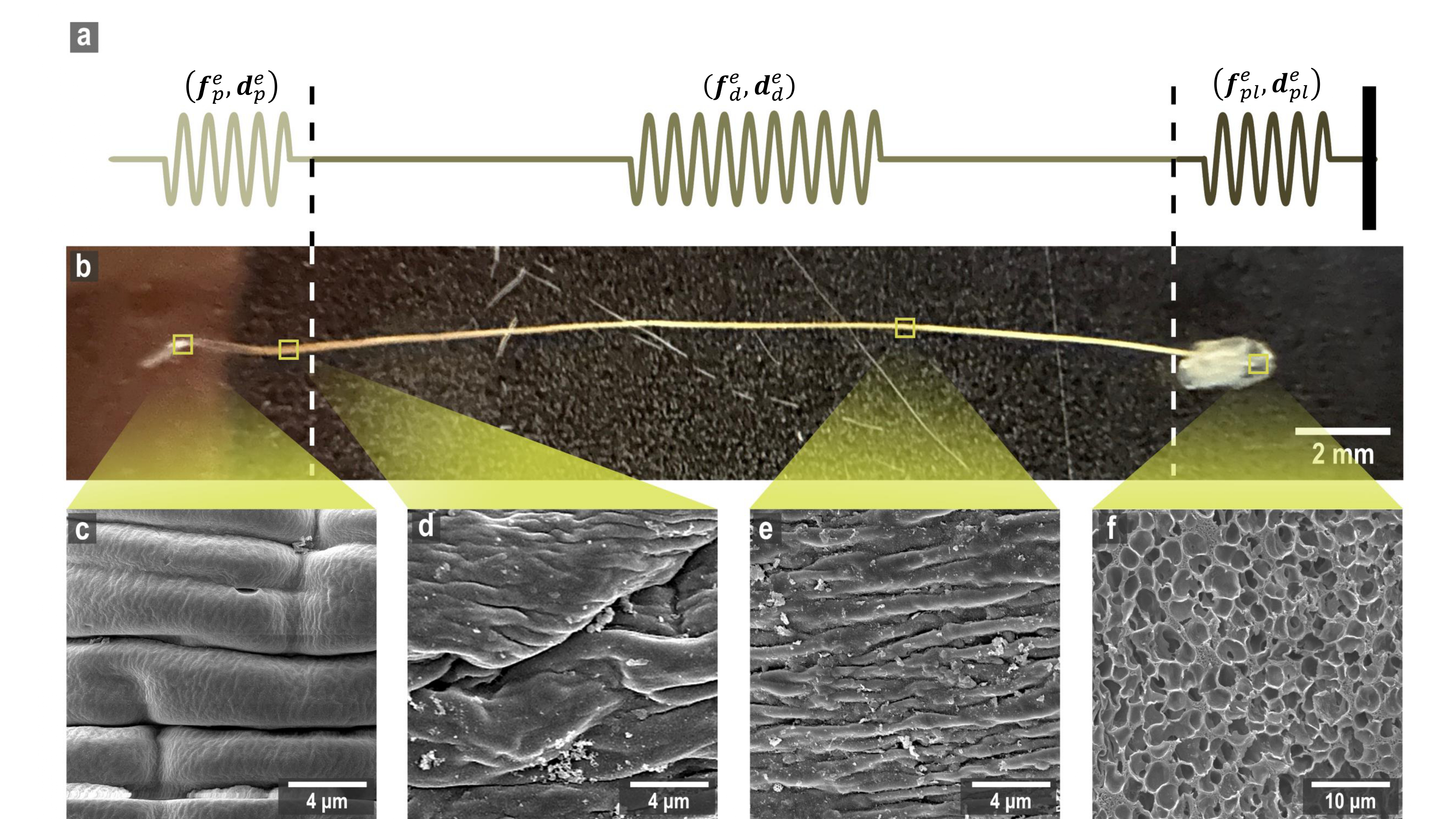}
 \caption{(a) The idealization of the thread-plaque mechanics model, (b) the overall thread-plaque structure, and (c)-(f) the micro-structures of proximal portion, transition portion, distal portion and plaque core, respectively. }
 \label{Fig7}
\end{figure*}

Four different microstructures from the mussel byssus system were identified by SEM analysis (Fig. \ref{Fig7}c-f). For the proximal portion, it can be seen that numerous folds of various sizes cover almost the entire surface of the proximal portion, and the furrows between the folds are parallel or perpendicular to the longitudinal axis (Fig. \ref{Fig7}c). The surface of the distal portion is smooth, but longitudinal patterns of approximately 1 \(\mu \mathrm{m}\) in width can still be seen on the surface (Fig. \ref{Fig7}e). Meanwhile, the surface of the transition part between these two portions has the characteristics of both, with a smoother surface and a reduced number of folds compared to the proximal portion, and longitudinal patterns begin to appear (Fig. \ref{Fig7}d). Due to the difficulty in identifying the location of the transition portion during the experiments, it was assumed that mussel threads were mainly composed of proximal and distal portions in this research. Furthermore, the mussel plaque with a porous core structure that is distinctly different from that of the thread is shown in Fig. \ref{Fig7}f. Different microstructural characteristics and protein composition\cite{coyne1997extensible,qin1997tough,wang2018recombinant} may be the reasons for the difference in material properties for different portions and provide evidence for the feasibility of the spring system assumption, which is the main basis of inverse approaches (Figs. \ref{Fig7}a and b).
\subsection{Micro-tensile test results}
\begin{figure*}
 \centering
 \includegraphics[height=10cm]{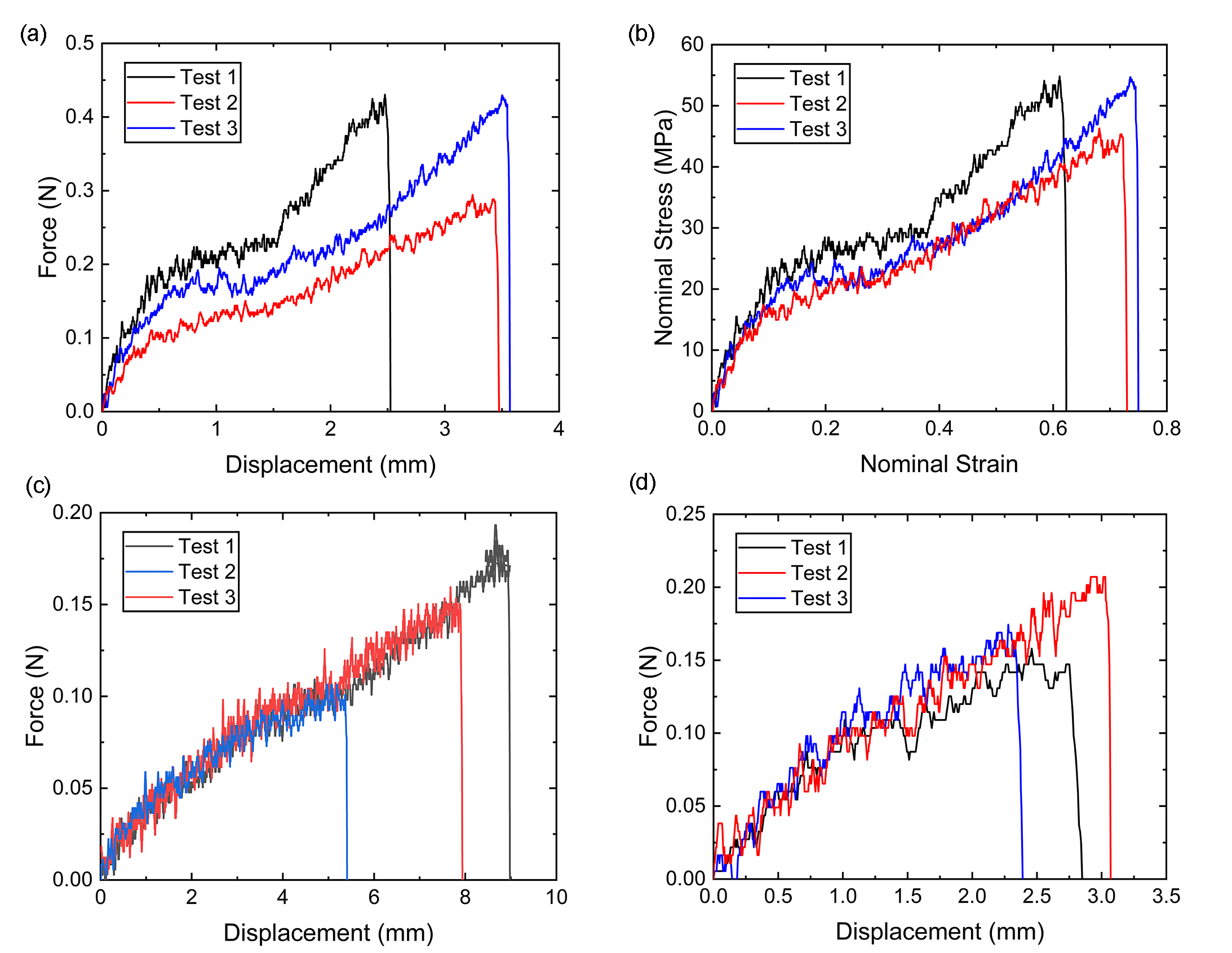}
 \caption{(a) The tested distal portion force-displacement curves, (b) the converted nominal stress-strain curves, (c) the tested whole thread (proximal and distal) portion force-displacement curves, and (d) the tested distal-plaque portion force-displacement curves.}
 \label{Fig8}
\end{figure*}

It is evident from Figs. \ref{Fig8}a and b that the force-displacement and strain-stress curves obtained from the tensile tests in the distal portions exhibit different failure points, reflecting the unique characteristics of each thread.\cite{price1981byssus,moeser2006seasonal,aldred2007tensile} Nevertheless, the distal portion tests show consistent failure mechanism patterns, suggesting the applicability of a single material constitutive model to represent the distal portion. 

Meanwhile, the force-displacement curves of the combined distal and proximal portions and the combined distal and plaque portions are shown in Figs. \ref{Fig8}c and d, respectively. An inflection point can be found in some of the force-displacement curves of the combined distal and plaque portions (Fig. \ref{Fig8}c), which may indicate that ultimate failure of the combined portions is likely to occur after the yield point of the distal portion. On the contrary, the force-displacement curves of the combined distal and plaque portions did not show an obvious inflection point, suggesting that plaque detachment may occur before the yield point of the distal section. The comparison of Figs. \ref{Fig8}c and d also demonstrates that the failure displacement of the combined distal and proximal portions is larger than the displacement of the combined distal and plaque portions, and the failure force varies from 0.1 N - 0.2 N during these tests. 
\subsection{Material properties of individual constituents}
\subsubsection{Material properties of the distal portion.~~}
\begin{figure}[h]
\centering
  \includegraphics[height=7cm]{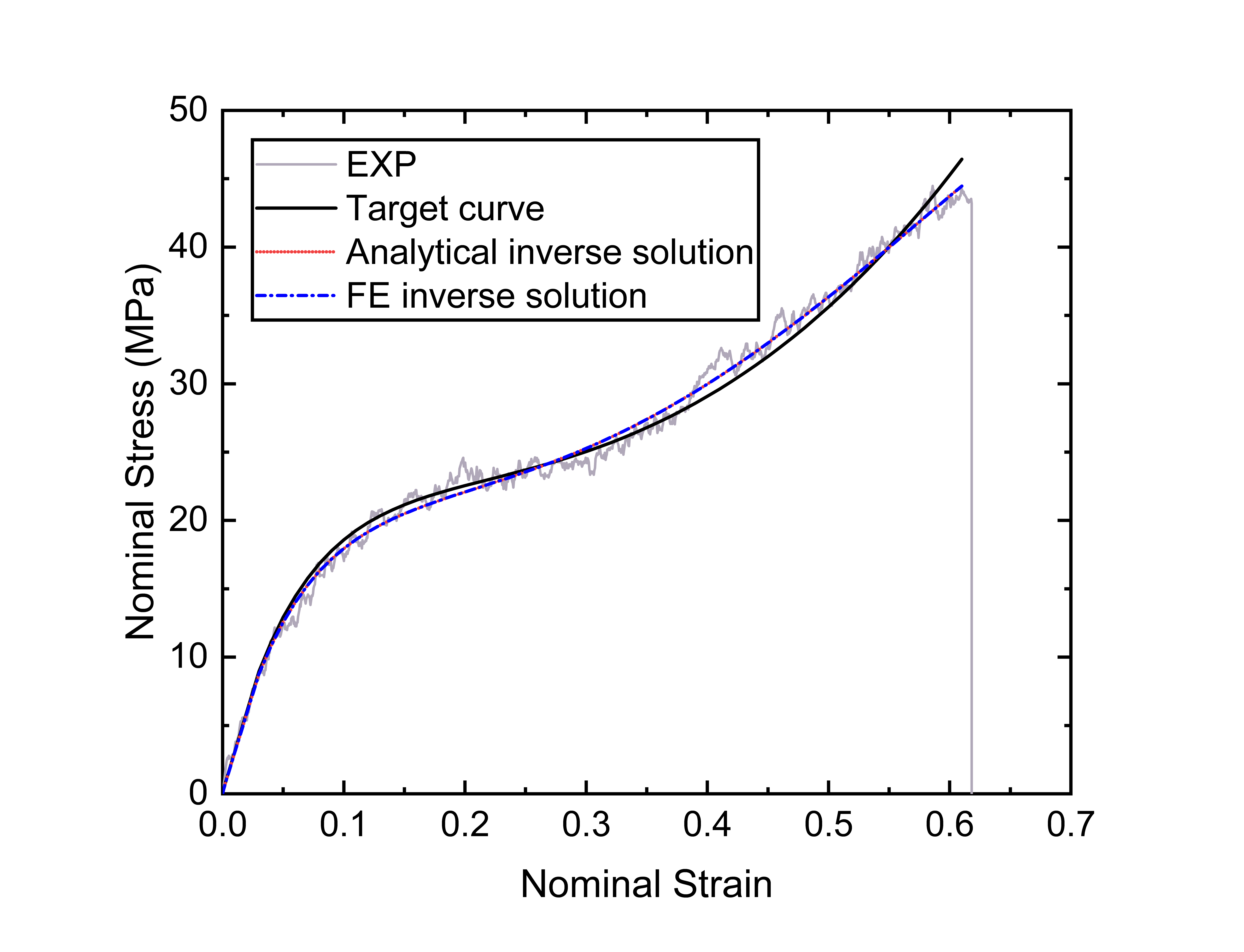}
  \caption{Nominal stress-strain curves of the distal portion obtained by experiment, fitting and inverse.}
  \label{Fig9}
\end{figure}

The "fminsearch" method in MATLAB was first used to generate a target curve that closely matched the experimental data, which was then used as input to the inverse approach to speed up the inverse calculation. The optimised curves of the analytical and FE inverse approaches with the assumed hyperelastic Odgen model show close agreement with the target curve (Fig. \ref{Fig9}). The material properties obtained from both inverse approaches are shown in Table 1 and were used as input for the latter proximal and plaque inverse approaches.
\begin{table}[h]
\small
\centering
  \caption{\ Material properties for the distal portion}
  \label{tbl:example1}
  \begin{tabular*}{0.48\textwidth}{@{\extracolsep{\fill}}lll}
    \hline
    Material constants & Analytical inverse solution & FE inverse solution \\
    \hline
    \(\mu_1 (MPa)\) & -529 & -528 \\
    \(\alpha_1\) & 7.19 & 6.69 \\
    \(\mu_2 (MPa)\) & 425 & 426 \\
    \(\alpha_2\) & 7.12 & 6.79 \\
    \(\mu_3 (MPa)\) & 229 & 229 \\
    \(\alpha_3\) & -14.5 & -12.4 \\
    \hline
  \end{tabular*}
\end{table}
\subsubsection{Material properties of the proximal portion.~~}
\begin{figure}[h]
\centering
  \includegraphics[height=7cm]{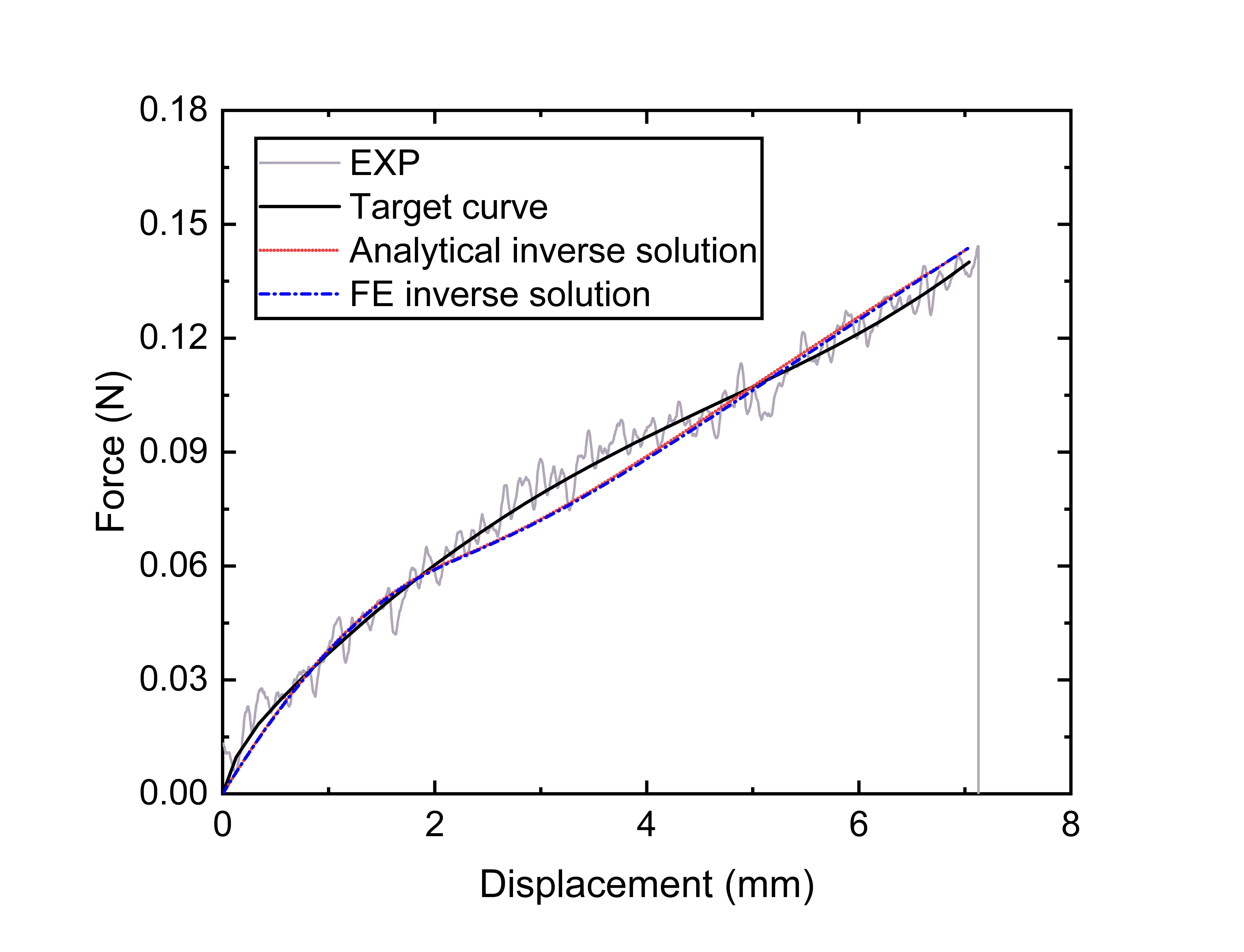}
  \caption{Nominal force-displacement curves of the thread (proximal and distal) portion obtained by experiment, fitting and inverse.}
  \label{Fig10}
\end{figure}

The process of optimising the material parameters for the proximal portion followed the method shown in Fig. \ref{Fig6}. Different thread dimensions can cause a significant variation in the force-displacement curve.  Therefore, instead of using the average experimental value, test 2 was selected from the three thread (proximal and distal) portion force-displacement curves in Fig. \ref{Fig8}a to remove the noise and set as the target curve. The force-displacement curves from inverse approaches on the basis of the assumed Neo-Hookean model are in good agreement with the experimental result (Fig. \ref{Fig10}). Therefore, the material parameter of the proximal portion \(C^{p}_1\) and residuals from both inverse approaches are given in Table 2. 
\begin{table}[h]
\small
\centering
  \caption{\ Proximal portion material parameter \(C^p_1 (MPa)\) and the min of the sum of square residuals \((N^2)\)}
  \label{tbl:example1}
  \begin{tabular*}{0.48\textwidth}{@{\extracolsep{\fill}}lll}
    \hline
    Inverse methods & \(C^p_1 (MPa)\) & \(\min\left\|\textbf{\textit{F}}\left(C^p_1\right)\right\|_2^2 (N^2)\)  \\
    \hline
    Analytical inverse method & 9.16 & 0.0004 \\
    FE-based inverse method & 11.2 & 0.0005 \\
    \hline
  \end{tabular*}
\end{table}
\subsubsection{Material properties of the plaque portion.~~}
\paragraph{Material parameter \(C^{p l}_1\) of the plaque portion}

\begin{figure*}
 \centering
 \includegraphics[width=1 \linewidth]{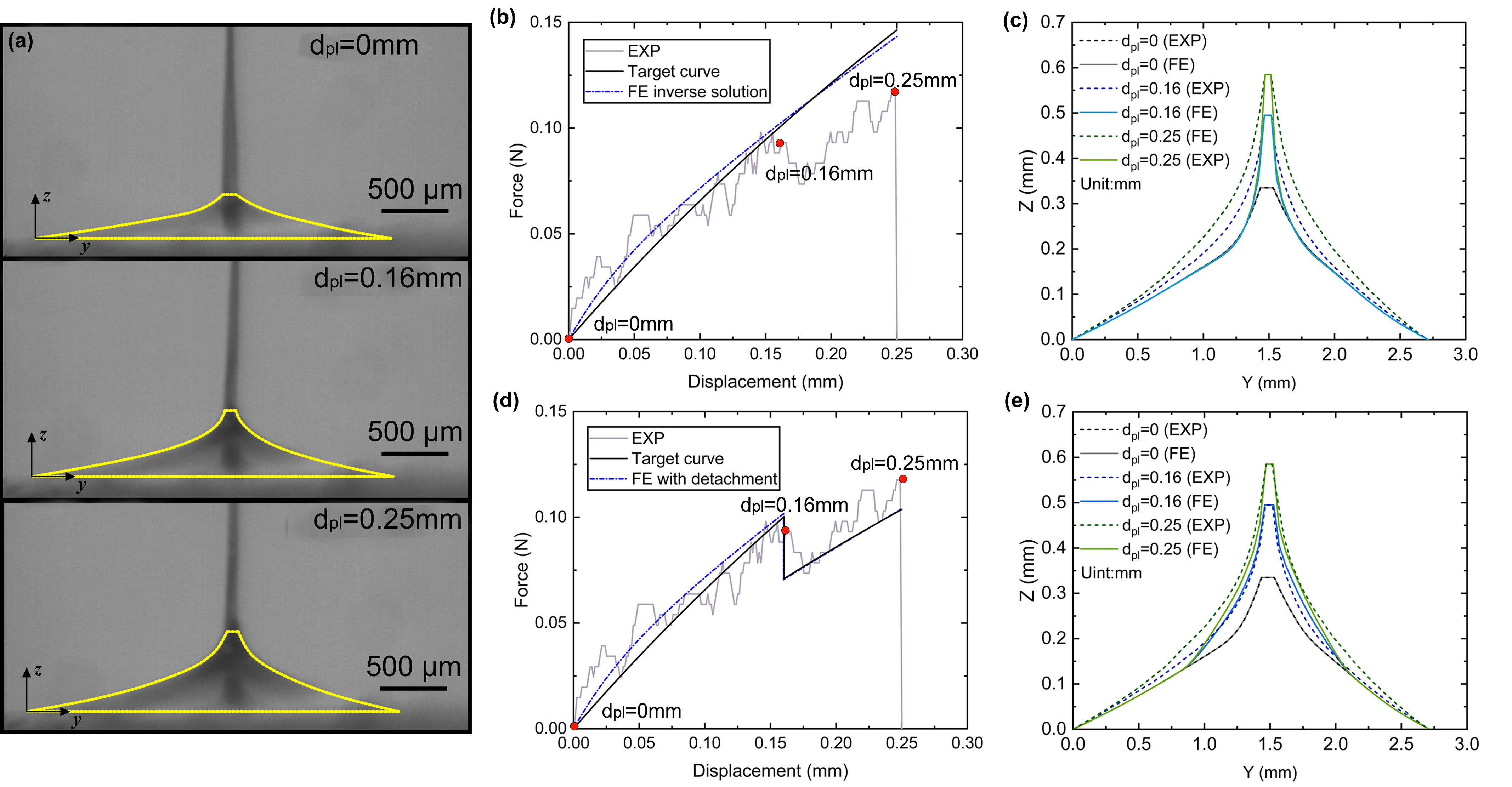}
 \caption{(a) The side view of plaque deformation at \(d_{p l}=0mm\), \(d_{p l}=0.16mm\), \(d_{p l}=0.25mm\). (b) The force-displacement curves without detachment at \(d_{p l}=0 mm, 0.16 mm, 0.25 mm\), (c) the deformed shape comparison of the plaque portion without detachment at \(d_{p l}=0 mm, 0.16 mm, 0.25 mm\), (d) the force-displacement curves with detachment at \(d_{p l}=0 mm, 0.16 mm, 0.25 mm\), and (e) the deformed shape comparison of the plaque portion with detachment at \(d_{p l}=0 mm, 0.16 mm, 0.25 mm\). }
 \label{Fig11}
\end{figure*}

Due to the complex and challenging nature of conducting tensile tests on the small and fragile mussel plaques, the distal-plaque tensile test results were analysed instead of conducting plaque-only experiments. From the experimental video, the displacement of the plaque portion prior to failure was measured to be 0.25 mm. Subsequently, by applying eqn (19) to the force data obtained from the distal-plaque test 1 (Fig. \ref{Fig8}d), the force-displacement curve for the individual plaque can be constructed, as shown in Figures. \ref{Fig11}b and d. Furthermore, three deformation shapes at different displacements were obtained from test 1 (Fig. \ref{Fig11}a). The leftmost points of the deformed shapes were set as the coordinate origins, and the shapes were further converted into three shape curves using Graph Grabber v 2.0.1 (Figs. \ref{Fig11}c and e).

The experimental force-displacement and shape curves (Figs. \ref{Fig11}b-e) were used as input for the FE inverse approach, then the material parameter for the plaque portion \(C^{p l}_1\) was output as 1.2 \(MPa\) with a minimum sum of squared residuals 0.137 \(mm^2\) (Table 3). The general agreement between the force-displacement curves and the shape curves obtained from the experimental results and the inverse results can be seen in Figs. \ref{Fig11}b and c. However, by carefully comparing the shape curves of the FE and the experiment, it can be observed that the shapes of the FE model with a fully fixed bottom surface showed obvious deformation only in the upper part (Fig. \ref{Fig11}c). On the other hand, the shapes of the experiment showed a more monolithic deformation, which also occurred in the area of the plaque that is far away from the stretch thread (Fig. \ref{Fig11}c). Meanwhile, after \(d_{p l}=0.16 mm\), the experimental curves were slightly lower than the simulated curves (Fig. \ref{Fig11}b). Therefore, the results could indicate a possible detachment on the bottom surface at \(d_{p l}=0.16 mm\) during the experiment. 
\begin{table}[h]
\small
\centering
  \caption{\ Plaque portion material parameter \(C^{p l}_1(MPa)\) and the minimum value of the sum of square residuals \((mm^2)\) from the FE inverse approach}
  \label{tbl:example1}
  \begin{tabular*}{0.48\textwidth}{@{\extracolsep{\fill}}lll}
    \hline
     & Inverse output without detachment  \\
    \hline
    \(C^{p l}_1\) & 1.2 \\
    \(\min\left[R\left(C_1^{p l}\right)\right]\)  & 0.137 \\
    \hline
  \end{tabular*}
\end{table}
\paragraph{Potential bottom detachment radius optimisation}

In order to further analyse the bottom surface detachment radius at \(d_{p l}=0.16 mm\), it was assumed that the detachment area was a circle and that the projection of the centre point on the upper surface to the lower surface was used as the detachment centre point. Furthermore, the material parameter \(C^{p l}_1\) was assumed to be the same as \(1.2 MPa\) and the detachment radius from \(d_{p l}=0.16 mm\) to \(d_{p l}=0.25 mm\) was assumed to be constant. The updated target force-displacement curve shown in Fig. \ref{Fig11}d was applied as the new input. Meanwhile, the radius of bottom surface detachment was optimised inside the FE inverse approach instead of the material parameter \(C^{p l}_1\). The 3D FE model used within this FE inverse approach was the same as in Section 2.3. The only difference was that the fixed boundary conditions at the bottom points within the detachment area were removed at \(d_{p l}=0.16 mm\) to simulate the bottom surface detachment during the optimisation process. Specifically, the detachment radius had to be less than the distance from the centre of the detachment to the outer boundary of the bottom surface. 

The updated inverse results with detachment can be seen in Figs. \ref{Fig11}d and e with a detachment radius = 0.6 \(mm\) and a less minimal sum of squared residuals 0.01 \(mm^2\) (Table 4). The FE and experiment deformed shapes at \(d_{p l}=0.16 mm\) and \(0.25 mm\) were more consistent than the previous results (Figs. \ref{Fig11}d and e). The discrepancy between the FE and the experimental deformed shapes at \(d_{p l}=0.25 mm\) suggests a possible detachment range greater than 0.6 \(mm\). However, it is important to note that 0.6 \(mm\) represents the maximum detachment radius without touching the edge of the bottom surface. This observation suggests that the detachment might be a dynamic process in which the detachment range gradually expands and is not static. Overall, the implementation of shape optimisation in the mussel plaque's FE inverse approach leads to a more accurate result. 
\begin{table}[h]
\small
\centering
  \caption{\ Bottom surface detachment radius \(r_b(mm)\) and the minimum value of the sum of square residuals \((mm^2)\) from the FE inverse approach}
  \label{tbl:example1}
  \begin{tabular*}{0.48\textwidth}{@{\extracolsep{\fill}}lll}
    \hline
     & FE inverse approach output \\
    \hline
    \(r_b\) & 0.6 \\
    \(\min\left[R\left(C_1^{p l}\right)\right]\) & 0.01 \\
    \hline
  \end{tabular*}
\end{table}

\section{Discussion}
In recent decades, several reports have investigated the tensile stress-strain behaviour of the mussel byssus system as a whole and its distal and proximal portions individually.\cite{carrington2004mechanical,bell1996mechanical,desmond2015dynamics,waite2002elastomeric} However, due to the size of the mussel byssus system and the complexity of the biomaterial composition, there are still some challenges to comprehensively understand its mechanism using experimental or numerical methods. In our study, the analytical and FE inverse approaches were applied for the first time to obtain the hyperelastic properties of the constituents of the mussel byssus system in this investigation, which could simplify the process of obtaining experimentally challenging material properties of the mussel byssus system or any other compound material system. The results show that the force-displacement curves based on the identified material properties from both inverse approaches are in good agreement with the micro-tensile test results, which demonstrates the feasibility of inverse methods for identifying the material properties of biomaterials. Notably, the biological characteristics of mussel plaques contribute to the variability in plaque underside shapes. Even plaques from the same mussel exhibit different structures depending on the temperature or PH levels. \cite{bernstein2020effects,filippidi2015microscopic} This investigation focuses on the application of the finite element (FE) inverse method, where the selected plaque had a nearly heart-shaped underside. However, in the study of Desmond et al. (2015) or Pang and Liu (2023), the undersides of the plaques showed an oval or square shape. \cite{desmond2015dynamics,pang2023quasi} Therefore, the integration of the inverse method with basic mechanical and optical tests is essential for a comprehensive analysis. Furthermore, it is pertinent to highlight that our experimental protocol only applied a stretching angle of $90\circ$ rather than aligning with the natural growth direction of mussels, as the research focus is to assess the feasibility of inverse methods to obtain the material properties of the mussel byssus system. Pang and Liu (2023) have conducted a detailed analysis to explain why the natural growth angle of mussel plaques is close to $15\circ$ using directional tensile experiments and FE simulations.\cite{pang2023quasi} Additionally, the deformed side shapes and the potential detachment of the plaque bottom surface were also included in the FE inverse approach to further increase the accuracy of the results. Specifically, in the study conducted by Waite et al. (2019), the spring constant and Young's modulus of the mussel plaque were obtained by simplifying the tress-strain curve of the combined distal and plaque system into three linear regions,\cite{mcmeeking2019force} while the consistency between our experimental and inverse curves suggests that the third-order Ogden model may be a more accurate option for characterising the material properties of distal threads, and the Neo-Hookean model may be a better choice for describing the material properties of both proximal threads and mussel plaques. 

In the present study, the tensile behaviour of the different portions of the mussel byssus system, as well as the microstructures of the different portions derived from micro-tensile tests, are also in agreement with the results of previous studies.\cite{carrington2004mechanical,bell1996mechanical,filippidi2015microscopic,vekhova2021byssal} The unique folds and furrows found in the proximal portion might be one of the reasons why the proximal portion is more ductile than the distal portion, as they allow the proximal portions more room to stretch and expand, helping to dissipate the strain energy (Figs. \ref{Fig7}c and e). Meanwhile, the variety in protein composition could be another reason to explain the differences in the tensile properties of the different portions. The biochemical composition of the mussel byssus system has been well researched, particularly the protein composition and metals.\cite{qin1997tough,coyne1997extensible,hwang2010protein,suhre2014structural,schmitt2015role} The distal portion is enriched in PreCol-D, whereas the proximal part is enriched in PreCol-P. PreCol-P with elastin-like folding sequences in the flanking domains is embedded in the matrix proteins of the proximal portion to provide greater elasticity. While preCol-D with spider silk-like sequences in the flanking domains could be reinforced with a \(\beta\)-sheet crystalline particle matrix to provide greater strength and toughness in the distal portion.\cite{golser2018routes} Additionally, the protein composition of the plaques differs from that of the threads, which is mainly composed of MFPs that help adhesion of the mussel plaque. The material properties, chemical composition, and microstructure of these three portions contribute to a better understanding of the mechanism and energy consumption underlying the composite characteristics of the mussel byssus system. 

Although identifying the material properties of each portion in the mussel byssus system is a significant first step in analysing the mechanism of how mussels remain stable on the seafloor against persistent waves, future investigation is needed once the material properties have been obtained. For example, mussel threads are a composite of an external cuticle and a fibrous core, whereas mussel plaques are a composite of a cuticle and a porous core. It is not yet clear how the cuticle layer and core affect tensile behaviour, respectively. Meanwhile, several questions have not been addressed, such as the transition mechanism between the proximal portion and the distal portion, the dynamic plaque detachment process within the inverse approach and the fracture properties of these different portions. In addition, the study by Wilhelm et al. (2017) found that due to various forms of microstructural damage, the force-displacement curves of the mussel byssus system under multiple loading cycles showed a hysteresis behaviour, but interestingly the material strength remained comparable to that of the monotonically loaded mussel byssus system.\cite{wilhelm2017influence} In our research, the mussel byssus system was assumed to be hyperelastic materials without considering any microstructural damage before the final failure. Therefore, how to further simulate and analyse the damage and self-healing of the mussel byssus systems under multiple loading cycles using inverse methods is also an interesting future research direction. 

In our study, we introduced analytical and FE inverse methods to obtain the material properties of the constituents of the mussel byssus system and validated the feasibility of these methods, which provide a new approach to analyse the composite behaviour of the mussel byssus system and any other biocomposite material for future biomaterial mechanism research. Meanwhile, the diversity of mussel byssus system mechanisms also provides many insights for mimetic applications in engineering and materials design, such as the design of anchor systems using the feature of the mussel byssus system to reduce fatigue in the mooring lines.  
\section{Conclusions}
In this research, both analytical and FE inverse methods have been applied for the first time to determine the individual material properties of the mussel plaque and the proximal portion. Those inverse methods were critical due to the inherent difficulty in obtaining these properties through conventional tests. The inverse methods utilised experimental force-displacement curves from different constituents of the mussel byssus system and side-deformed shapes of mussel plaques as optimisation metrics.

The findings indicate that these inverse techniques effectively identify the individual material properties of the proximal thread and the plaque using reasonably assumed constitutive material models with the sum of square residuals equal to 0.0004 (\(N^2\)) and 0.01 (\(mm^2\)), respectively. Meanwhile, the results support the accurate representation of the material properties of the distal portion with the Ogden hyperelastic model and the proximal and plaque portions with the Neo-Hookean hyperelastic model. Additionally, this study has also conducted a comprehensive characterisation of the microstructure and material properties of the three different portions within the mussel byssus system. The distal portion exhibits elevated strength but reduced ductility in contrast to the proximal portion. The variations in the tensile behaviour among these portions could be attributed to differences in their microstructure or chemical composition. Finally, the study has also discussed the application of the FE inverse method for the calculation of the potential plaque bottom surface detachment radius. 

The material parameters obtained from the inverse methods offer an opportunity for in-depth analysis of the material behaviour of the mussel byssus system. Future studies could focus on further improving the precision of inverse approaches by assessing the transition mechanisms between different portions and monitoring the dynamic detachment processes of the plaque bottom surface. Additionally, it will be necessary to determine the fracture characteristics of these three portions in the future and simulate the entire fracture tensile process using a comprehensive 3D finite element model of the mussel byssus system. 

\section*{Conflicts of interest}
There are no conflicts to declare.

\section*{Acknowledgements}
This work was funded by the Leverhulme Trust Research Grant Scheme, UK (No. RPG-2020-235) and NanoCAT Research Grant, the University of Nottingham.

\biboptions{numbers,sort&compress}
\bibliographystyle{elsarticle-num}
\bibliography{References} 

\begin{thebibliography}{10}
\expandafter\ifx\csname url\endcsname\relax
  \def\url#1{\texttt{#1}}\fi
\expandafter\ifx\csname urlprefix\endcsname\relax\def\urlprefix{URL }\fi
\expandafter\ifx\csname href\endcsname\relax
  \def\href#1#2{#2} \def\path#1{#1}\fi

\bibitem{lee2011mussel}
B.~P. Lee, P.~B. Messersmith, J.~N. Israelachvili, J.~H. Waite, Mussel-inspired
  adhesives and coatings, Annual review of materials research 41 (2011)
  99--132.

\bibitem{perepelizin2011thermal}
P.~V. Perepelizin, D.~Boltovskoy, Thermal tolerance of limnoperna fortunei to
  gradual temperature increase and its applications for biofouling control in
  industrial and power plants, Biofouling 27~(6) (2011) 667--674.

\bibitem{witman1984mussels}
J.~D. Witman, T.~H. Suchanek, Mussels in flow: drag and dislodgement by
  epizoans, Marine Ecology Progress Series 16~(3) (1984) 259--268.

\bibitem{li2019molecular}
Y.~Li, Y.~Cao, The molecular mechanisms underlying mussel adhesion, Nanoscale
  Advances 1~(11) (2019) 4246--4257.

\bibitem{lin2007adhesion}
Q.~Lin, D.~Gourdon, C.~Sun, N.~Holten-Andersen, T.~H. Anderson, J.~H. Waite,
  J.~N. Israelachvili, Adhesion mechanisms of the mussel foot proteins mfp-1
  and mfp-3, Proceedings of the National Academy of Sciences 104~(10) (2007)
  3782--3786.

\bibitem{hwang2010protein}
D.~S. Hwang, H.~Zeng, A.~Masic, M.~J. Harrington, J.~N. Israelachvili, J.~H.
  Waite, Protein-and metal-dependent interactions of a prominent protein in
  mussel adhesive plaques, Journal of biological chemistry 285~(33) (2010)
  25850--25858.

\bibitem{allen1976observations}
J.~Allen, M.~Cook, D.~Jackson, S.~Preston, E.~Worth, Observations on the rate
  of production and mechanical properties of the byssus threads of mytilus
  edulis l, Journal of Molluscan Studies 42~(2) (1976) 279--289.

\bibitem{crisp1985adhesion}
D.~Crisp, G.~Walker, G.~Young, A.~Yule, Adhesion and substrate choice in
  mussels and barnacles, Journal of Colloid and Interface Science 104~(1)
  (1985) 40--50.

\bibitem{bell1996mechanical}
E.~C. Bell, J.~M. Gosline, Mechanical design of mussel byssus: material yield
  enhances attachment strength, Journal of Experimental Biology 199~(4) (1996)
  1005--1017.

\bibitem{carrington2004mechanical}
E.~Carrington, J.~M. Gosline, Mechanical design of mussell byssus: Load cycle
  and strain rate dependence (2004).

\bibitem{burkett2009method}
J.~R. Burkett, J.~L. Wojtas, J.~L. Cloud, J.~J. Wilker, A method for measuring
  the adhesion strength of marine mussels, The Journal of Adhesion 85~(9)
  (2009) 601--615.

\bibitem{desmond2015dynamics}
K.~W. Desmond, N.~A. Zacchia, J.~H. Waite, M.~T. Valentine, Dynamics of mussel
  plaque detachment, Soft matter 11~(34) (2015) 6832--6839.

\bibitem{pang2023quasi}
Y.~Pang, T.~Liu, Quasi-static responses of marine mussel plaques attached to
  deformable wet substrates under directional tensions, arXiv preprint
  arXiv:2305.15129 (2023).

\bibitem{holten2008mussel}
N.~Holten-Andersen, J.~Waite, Mussel-designed protective coatings for compliant
  substrates, Journal of dental research 87~(8) (2008) 701--709.

\bibitem{waite1998peculiar}
J.~H. Waite, X.-X. Qin, K.~J. Coyne, The peculiar collagens of mussel byssus,
  Matrix biology 17~(2) (1998) 93--106.

\bibitem{gosline2002elastic}
J.~Gosline, M.~Lillie, E.~Carrington, P.~Guerette, C.~Ortlepp, K.~Savage,
  Elastic proteins: biological roles and mechanical properties, Philosophical
  Transactions of the Royal Society of London. Series B: Biological Sciences
  357~(1418) (2002) 121--132.

\bibitem{hagenau2009structural}
A.~Hagenau, H.~A. Scheidt, L.~Serpell, D.~Huster, T.~Scheibel, Structural
  analysis of proteinaceous components in byssal threads of the mussel mytilus
  galloprovincialis, Macromolecular bioscience 9~(2) (2009) 162--168.

\bibitem{areyano2022viscoelastic}
M.~Areyano, E.~Valois, I.~Sanchez~Carvajal, I.~Rajkovic, W.~R. Wonderly,
  A.~Kossa, R.~M. McMeeking, J.~H. Waite, Viscoelastic analysis of mussel
  threads reveals energy dissipative mechanisms, Journal of the Royal Society
  Interface 19~(188) (2022) 20210828.

\bibitem{harrington2009collagen}
M.~J. Harrington, H.~S. Gupta, P.~Fratzl, J.~H. Waite, Collagen insulated from
  tensile damage by domains that unfold reversibly: In situ x-ray investigation
  of mechanical yield and damage repair in the mussel byssus, Journal of
  structural biology 167~(1) (2009) 47--54.

\bibitem{sun2001oxidative}
C.~Sun, E.~Vaccaro, J.~H. Waite, Oxidative stress and the mechanical properties
  of naturally occurring chimeric collagen-containing fibers, Biophysical
  Journal 81~(6) (2001) 3590--3595.

\bibitem{harrington2007holdfast}
M.~J. Harrington, J.~H. Waite, Holdfast heroics: comparing the molecular and
  mechanical properties of mytilus californianus byssal threads, Journal of
  Experimental Biology 210~(24) (2007) 4307--4318.

\bibitem{xu2019metal}
Q.~Xu, M.~Xu, C.-Y. Lin, Q.~Zhao, R.~Zhang, X.~Dong, Y.~Zhang, S.~Tian,
  Y.~Tian, Z.~Xia, Metal coordination-mediated functional grading and
  self-healing in mussel byssus cuticle, Advanced Science 6~(23) (2019)
  1902043.

\bibitem{zechel2019healing}
S.~Zechel, M.~D. Hager, T.~Priemel, M.~J. Harrington, Healing through
  histidine: Bioinspired pathways to self-healing polymers via imidazole--metal
  coordination, Biomimetics 4~(1) (2019) 20.

\bibitem{aldred2007tensile}
N.~Aldred, T.~Wills, D.~Williams, A.~Clare, Tensile and dynamic mechanical
  analysis of the distal portion of mussel (mytilus edulis) byssal threads,
  Journal of the Royal Society Interface 4~(17) (2007) 1159--1167.

\bibitem{smeathers1979mechanical}
J.~Smeathers, J.~Vincent, Mechanical properties of mussel byssus threads,
  Journal of Molluscan Studies 45~(2) (1979) 219--230.

\bibitem{mcmeeking2019force}
N.~Cohen, J.~H. Waite, R.~M. McMeeking, M.~T. Valentine, Force distribution and
  multiscale mechanics in the mussel byssus, Philosophical Transactions of the
  Royal Society B 374~(1784) (2019) 20190202.

\bibitem{vaccaro2001yield}
E.~Vaccaro, J.~H. Waite, Yield and post-yield behavior of mussel byssal thread:
  a self-healing biomolecular material, Biomacromolecules 2~(3) (2001)
  906--911.

\bibitem{filippidi2015microscopic}
E.~Filippidi, D.~G. DeMartini, P.~Malo~de Molina, E.~W. Danner, J.~Kim, M.~E.
  Helgeson, J.~H. Waite, M.~T. Valentine, The microscopic network structure of
  mussel (mytilus) adhesive plaques, Journal of The Royal Society Interface
  12~(113) (2015) 20150827.

\bibitem{guide1998mathworks}
M.~U. Guide, The mathworks, Inc., Natick, MA 5~(333) (1998) 4.

\bibitem{waite2002elastomeric}
J.~H. Waite, E.~Vaccaro, C.~Sun, J.~M. Lucas, Elastomeric gradients: a hedge
  against stress concentration in marine holdfasts?, Philosophical Transactions
  of the Royal Society of London. Series B: Biological Sciences 357~(1418)
  (2002) 143--153.

\bibitem{ogden1972large}
R.~W. Ogden, Large deformation isotropic elasticity--on the correlation of
  theory and experiment for incompressible rubberlike solids, Proceedings of
  the Royal Society of London. A. Mathematical and Physical Sciences 326~(1567)
  (1972) 565--584.

\bibitem{rivlin1948large}
R.~S. Rivlin, Large elastic deformations of isotropic materials iv. further
  developments of the general theory, Philosophical transactions of the royal
  society of London. Series A, Mathematical and physical sciences 241~(835)
  (1948) 379--397.

\bibitem{Schnur}
D.~S. Schnur, N.~Zabaras, An inverse method for determining elastic material
  properties and a material interface, International Journal for Numerical
  Methods in Engineering 33~(10) (1992) 2039--2057.

\bibitem{sun2022overview}
W.~Sun, Z.~Yue, Z.~Wen, M.~Li, An overview on material parameter inverse and
  its application to miniaturized testing at elevated temperature, Journal of
  Materials Research and Technology (2022).

\bibitem{coyne1997extensible}
K.~J. Coyne, X.-X. Qin, J.~H. Waite, Extensible collagen in mussel byssus: a
  natural block copolymer, Science 277~(5333) (1997) 1830--1832.

\bibitem{qin1997tough}
X.-X. Qin, K.~J. Coyne, J.~H. Waite, Tough tendons: mussel byssus has collagen
  with silk-like domains, Journal of Biological Chemistry 272~(51) (1997)
  32623--32627.

\bibitem{wang2018recombinant}
J.~Wang, T.~Scheibel, Recombinant production of mussel byssus inspired
  proteins, Biotechnology Journal 13~(12) (2018) 1800146.

\bibitem{price1981byssus}
H.~A. Price, Byssus thread strength in the mussel, mytilus edulis, Journal of
  Zoology 194~(2) (1981) 245--255.

\bibitem{moeser2006seasonal}
G.~M. Moeser, E.~Carrington, Seasonal variation in mussel byssal thread
  mechanics, Journal of Experimental Biology 209~(10) (2006) 1996--2003.

\bibitem{bernstein2020effects}
J.~H. Bernstein, E.~Filippidi, J.~H. Waite, M.~T. Valentine, Effects of sea
  water ph on marine mussel plaque maturation, Soft Matter 16~(40) (2020)
  9339--9346.

\bibitem{vekhova2021byssal}
E.~Vekhova, The byssal apparatus in the pacific mussel, mytilus trossulus
  (bivalvia, mytilidae), from the sea of japan, Biology Bulletin 48~(9) (2021)
  1443--1451.

\bibitem{suhre2014structural}
M.~H. Suhre, M.~Gertz, C.~Steegborn, T.~Scheibel, Structural and functional
  features of a collagen-binding matrix protein from the mussel byssus, Nature
  communications 5~(1) (2014) 3392.

\bibitem{schmitt2015role}
C.~N. Schmitt, Y.~Politi, A.~Reinecke, M.~J. Harrington, Role of sacrificial
  protein--metal bond exchange in mussel byssal thread self-healing,
  Biomacromolecules 16~(9) (2015) 2852--2861.

\bibitem{golser2018routes}
A.~V. Golser, T.~Scheibel, Routes towards novel collagen-like biomaterials,
  Fibers 6~(2) (2018) 21.

\bibitem{wilhelm2017influence}
M.~H. Wilhelm, E.~Filippidi, J.~H. Waite, M.~T. Valentine, Influence of
  multi-cycle loading on the structure and mechanics of marine mussel plaques,
  Soft Matter 13~(40) (2017) 7381--7388.

\end{thebibliography}
\end{document}